\documentclass[preprint, aps, prb, amsmath, amssymb, showpacs, superscriptaddress, seceqn]{revtex4-1}
\usepackage{amsfonts, hyperref, times}

\usepackage{graphics}
\usepackage{bbold}
\usepackage{siunitx}

\usepackage{ragged2e}
\usepackage{subcaption}
\usepackage[demo]{graphicx}

\providecommand{\vect}[1]{{\boldsymbol{#1}}}

\begin{document}
  
\title{Potential implementation of Reservoir Computing models based on magnetic skyrmions}

\author{George Bourianoff}
\email[George Bourianoff is the corresponding author for discussions related to Reservoir Computing: ]{george.bourianoff@intel.com}
\affiliation{\footnotesize Intel Corporation, 1300 S. MoPac Exp, Austin, TX, 78746, U.S.}

\author{Daniele Pinna}
\affiliation{\footnotesize Institute of Physics, Johannes Gutenberg University Mainz, 55099 Mainz, Germany}

\author{Matthias Sitte}
\affiliation{\footnotesize Institute of Physics, Johannes Gutenberg University Mainz, 55099 Mainz, Germany}

\author{Karin Everschor-Sitte}
\affiliation{\footnotesize Institute of Physics, Johannes Gutenberg University Mainz, 55099 Mainz, Germany}
\email[Karin Everschor-Sitte is the corresponding author for discussions related to skyrmion physics: ]{kaeversc@uni-mainz.de}

\date{\today}

\begin{abstract}

Reservoir Computing is a type  of recursive neural network commonly used for recognizing and predicting spatio-temporal events relying on a complex hierarchy of nested feedback loops to generate a memory functionality. The Reservoir Computing paradigm does not require any knowledge of the reservoir topology or node weights for training purposes and can therefore utilize naturally existing networks formed by a wide variety of physical processes. Most efforts prior to this have focused on utilizing memristor techniques to implement recursive neural networks. This paper examines the potential of skyrmion fabrics formed in magnets with broken inversion symmetry that may provide an attractive physical instantiation for Reservoir Computing.

\end{abstract}


\maketitle

\section{Introduction}

A great deal has been written about the end of CMOS scaling, continuation of Moore's Law and the need for alternative models of computing and related technologies.   One of the most authoritative discussions on Moore's Law can be found in the ``final'' International Technology Roadmap for Semiconductors (ITRS)\cite{ITRS2015} published in 2015 and which had been continuously published since 1991. It predicted that CMOS transistors would quit shrinking in 2021 with the $\SI{5}{\nano\metre}$ node and that a great many technical challenges would need to be met for the $\SI{5}{\nano\metre}$ node to be economically viable.  Those technical challenges were primarily related to controlling the economic costs associated with lithography, packaging, testing and the process technology itself.  The technical challenges of gate leakage, interconnect power losses and material integration were all considered daunting and drove the economic issues to the point where further scaling was possible but not economically viable.  
The inescapable conclusion of the final ITRS document is that the scientific research community in collaboration with industry must investigate alternative models of computing including those which appear to be quite radical.

The universe of alternative models of computing is enormous and growing.  A nonexclusive list includes\cite{Adamatzky2017} membrane computing, DNA computing, immune computing, quantum computing, neuromorphic computing, in-materio computing,  swarm computing, analog computing, chaos / edge-of-chaos computing, computational aspects of dynamics of complex systems, self-organizing systems (e.g.\ multiagent systems, cellular automata, artificial life), and many others. 
In evaluating this list of alternative computational paradigms, one is reminded of Kroemer's Law\cite{Maiti2008} which states that ``the principal applications of any new and innovative technology always have been and will continue to be created by that new technology''.
One should therefore not only judge new technologies by how they fit in with present applications but specifically by their potential to create innovative applications themselves. Optimally, however, they will satisfy both criteria.  

Predicted\cite{Bogdanov1989} and subsequently discovered\cite{Muhlbauer2009a} over the past two decades, skyrmions are widely regarded as promising candidates for spintronic applications due to their room-temperature stability
\cite{Yu2011b, Yu2012, Gilbert2015,Woo2016,Boulle2016,Tomasello2017a} 
and mobility at ultra-low current densities\cite{Jonietz2010,Schulz2012}. Based on these alone, skrymions have been proposed for enhancing a spectrum of existing technologies such as racetrack memories\cite{Fert2013, Tomasello2014, Zhang2015c, Muller2016}, transistors\cite{Zhang2015g} and logic gates.\cite{Zhou2014, Zhang2015l} 
However, the integration of their intrinsic two-dimensional nature also enables their use for radically new technologies\cite{Pinna2017, Prychynenko2017, He2017, He2017a, Li2017a}. 

In this paper we focus on Reservoir Computing (RC) models\cite{Lukosevicius2009, Lukosevicius2012a, Burger2015, Lukosevicius2012, Verstraeten2007, Maass2002} implemented with self-organizing neural networks in complex magnetic textures.\cite{Prychynenko2017}
The nodes are represented by magnetic skyrmions and the random connectivity by low magnetoresistive pathways in the material. 
Specifically, we will consider the effects of anisotropic magnetoresistance (AMR) on the conductivity pathways in systems with broken 
bulk and surface inversion symmetry.
For purposes of placement in the taxonomy above, RC models are one category of neuromorphic computing and utilizing complex magnetic systems is an example of in-materio computing.\cite{Dale2017}  

The paper is organised as follows.  Sec.~\ref{sec:RC} contains a selective review of RC focusing on the requirements placed on the physical implementation system.  In Sec.~\ref{sec:skrymions} we first briefly review the skyrmion and micromagnetic literature relevant to RC implementation and then present new simulation results to address the suitability of magnetic skyrmion networks for RC.
In Sec.~\ref{sec:conclusions} we conclude how well the capabilities of magnetic substrates meet the needs of RC and delineate areas of future research.

\section{Reservoir Computing}
\label{sec:RC}

A Recurrent Neural Network (RNN) is a network of nonlinear processing units (similar to neurons in the brain) with weighted connections between them (synapses) characterized by a flow of information that feeds back in loops.  
These loops imbue the system with a notion of memory where an ``echo'' of previous input sequences persists over time. Generally, RNN's are trained to perform specific tasks by algorithmically tuning the weights of each connection in the network. However, the presence of feedback loops makes RNN extremely difficult to train because of bifurcation points and the tendency to exhibit chaotic solutions. As such, the same property responsible for their power is also responsible for their limited applications to date.

Reservoir Computing (RC) models address this problem by treating the recurrent part (the reservoir) differently than both the read-in and read-outs from it~\cite{Lukosevicius2012a}. This eliminates the need to train the large complex reservoir and only train the output weights. Because the latter are taken to not have any loops, they can be trained by straightforward linear regression techniques, i.e.\ least square algorithms. The usefulness of this framework derives from the reservoir's capacity to project any applied signals into a sparse, high-dimensional space where recognition becomes easier. This has been shown with mathematical rigor\cite{Maass2002, Verstraeten2007, Lukosevicius2012}, allowing RC to be successfully demonstrated for temporal signal recognition and prediction. 

Unlike designed computation, where each device has a specific role, computation in RC networks does not rely on specific devices in specific roles, but is encoded in the collective nonlinear dynamics excited by an applied input signal. This most attractive aspect of RC implies that it is unnecessary to have any knowledge related to the structure of the reservoir itself. It is not necessary to know the reservoir's structure, individual node connections, weights, nor any of their nonlinear characteristics. For this reason RC methods may use ``found'' networks that may be formed naturally by a wide variety of physical processes.

Reservoir Computing is typically further divided into two categories, Echo State Networks (ESNs) and Liquid State Networks (LSNs)\cite{Lukosevicius2012}. LSNs are characterized by nodes whose state is considered a continuous-time binary value (on/off) whose spiking frequency is determined by the activity of neighboring nodes. Even though they are not actively considered in this work, we note in passing that they are considered more reminiscent of bio-inspired neuromorphic computing models since they directly emulate the spiking potential generation and propagation observed in biological systems. On the other hand, ESNs are characterized by nodes whose state is defined by continuous values to be updated in discrete time steps depending on the state of nearby nodes. The ESN and LSN approaches alike both provide a similar function -- they project the input data into a high-dimensional space defined by the {\it state} of the entire reservoir.

\subsection{Echo State Networks}

\begin{figure}
\includegraphics[width=0.5\textwidth]{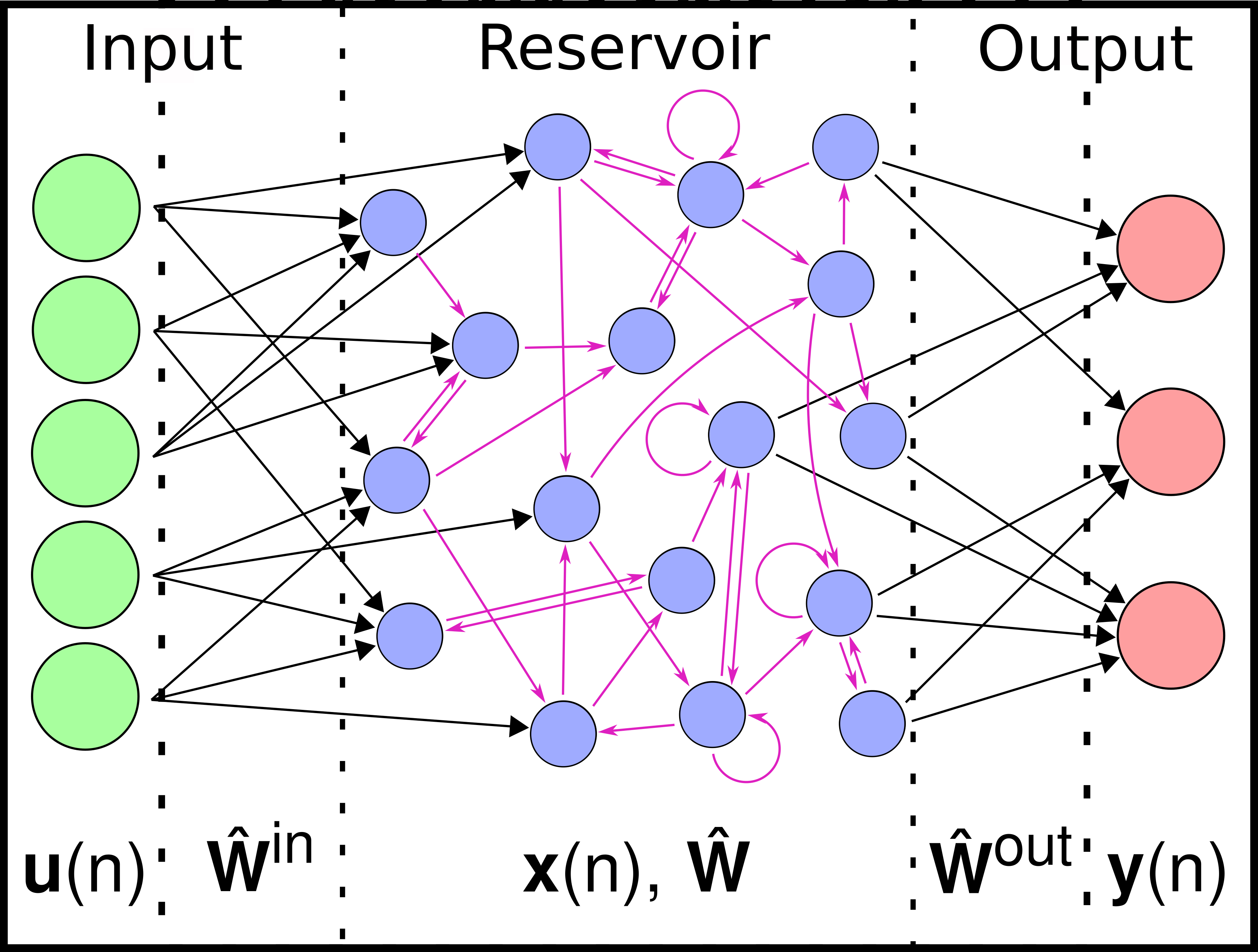}
\caption{Sketch of a generic echo state network. The nodes of the network are represented by dots where the color indicates their functionality as input (green) output (red) or internal network nodes (blue). The arrows represent the network connectivity, where black arrows are feedforward only and pink arrows might occur in birectionally.
}
\label{fig:echostatenetwork}
\end{figure}

A generic ESN is shown schematically in Fig.~\ref{fig:echostatenetwork}. 
It consists of an input layer, the network, and the output layer.
The individual nodes in the network are indicated by light blue dots and the nodes of the input (output) layer are represented by light green (red) dots.
 The arrows indicate connection paths. The input and output nodes represent a feedforward network (black arrows) whereas the reservoir nodes are in general bidirectional forming both single node and multimode loops (pink arrows). The combination of feedforward and feedback connections results in recursive operation. 

Following standard nomenclature\cite{Lukosevicius2012book}, the input signal $\vect u(n)$ is a discrete function of time, $\vect x(n)$ represents the state of the reservoir, e.g.\ the electric potential at every node, and $\vect y(n)$ represents the output signal. 
 The corresponding weighting functions are denoted as $\hat{\vect W}^{\mathrm{in}}$, $\hat{\vect W}$ and $\hat{\vect W}^{\mathrm{out}}$ for the input, the reservoir, and the output. In general, the weights of the reservoir and the input, i.e.\ $\hat{\vect W^{\mathrm{in}}}$ and $\hat{\vect W}$ are time-independent. As already stated, the reservoir weighting function $\hat{\vect W}$ can be unknown which allows ``found'' reservoirs like the random skyrmion networks discussed in this paper to be used.

The evolution of the reservoir state at time step $n$ in reaction to its input ${\vect u}(n)$ and the reservoir state at time step $n-1$ is defined as:
\begin{align}
\tilde{\vect x}(n)&=\mathrm{sig}\left(
\hat{\vect W}_{in} \cdot \vect u (n)  \right) + \hat{\vect W} \cdot\vect x(n-1),\label{eq:xprime} \\
\vect x(n)&= \lambda \, \vect x(n-1)+(1- \lambda)\, \tilde{\vect x}(n),
\label{eq:Xn}\\
\vect y(n)&= \hat{\vect W}_{\mathrm{out}} \cdot \vect{x}(n), 
\label{eq:Yn}
\end{align}
where $\tilde{\vect x}$ is a provisional state variable, $\mathrm{sig}(\cdot)$ is generically a sigmoidal function biasing the input signal so that the reservoir is excited but not saturated, and $\lambda$ is a {\it leakage} parameter characterizing the lossiness of the network's memory. Note, that for systems without leakage ($\lambda =0$), one has $\vect x(n) =\tilde{\vect x}(n)$. In a natural reservoir system, the specific form of the sigmoidal function is determined by the system's physical properties. 

The ultimate goal of the ESN is to classify similar input signals into identical outputs. This is achieved by first training the network on a sample set of pre-classified inputs $\vect u^{\mathrm{train}}(n)$. Denoting by $\vect y^{\mathrm{train}}(n)$ and $\vect y^{\mathrm{target}}(n)$ the system's response to the training inputs and the desired target output respectively, we can calculate the error $E$ between them, averaged over all $N_y$ output nodes and $T$ time steps:
\begin{equation}
\label{eq:Ey}
E(\vect{y}^{\mathrm{train}},\vect{y}^{\mathrm{target}})=\frac{1}{N_y}\sum_{i=1}^{N_y}\sqrt{\frac{1}{T}\sum_{n=1}^T(y^{\mathrm{train}}_i(n)-y_i^{\mathrm{target}}(n))^2}.
\end{equation}
The training challenge is defined by finding the one-dimensional scalar array of output weights $\hat{\vect W}^{\mathrm{out}}$ that minimizes the error function $E$. While the specific algorithms implementing the minimization task are beyond the scope of this paper, we would like to emphasize again that Eq.~\eqref{eq:Ey} only requires the output weights to be modified. This is inherently more efficient and robust than the methods required for training full RNNs. Furthermore, since training of the output weights does not modify the reservoir in any way, different features of the reservoir can be searched for simultaneously by setting multiple output arrays in parallel. This makes RC well suited for sensor fusion type applications.\cite{Palumbo2013, Konkoli2016}
	
To summarize, the reservoir requires a few qualitative key characteristics to properly function in an Echo State computing system:

\begin{itemize}
\item It must have a short term memory, i.e.\ be recursively connected and/or use nodes with internal memory. As discussed, this guarantees the reservoir's sensitivity to the input's temporal correlations. 

\item The dimensionality of the reservoir's state space must be much larger than the input array. This corresponds basically to the number of nodes in the reservoir. The larger the reservoir, the greater the separation and probability of the linear classifier being able to successfully recognise specific events.

\item The response of the reservoir must be a nonlinear function of its inputs and previous states. The stronger the nonlinearity effects are, the faster different input signals are spread in the reservoir's phase space facilitating classification during training. 

\item The reservoir's echo state time, defined as the timescale beyond which the reservoir dynamics effectively lose all initialization information, must be much larger than the largest relevant temporal correlations in the input signals. The reservoir's echo state time can be tuned by varying the leakage parameter $\lambda$.
\end{itemize}

These reservoir properties (and the parameters controlling them) determine the performance of the recursive network for spatial temporal event recognition and prediction. For computational materials like the magnetic textures considered in this paper, these (or an equivalent set) must be ultimately deduced by experimental measurements and tuned for optimal performance by leveraging physical insight over the materials being used.  
In the following section, we will proceed by reviewing the literature of magnetic skyrmions relevant to a potential RC implementation.

\section{Magnetic skyrmions and "skyrmion fabrics"}
\label{sec:skrymions}

Magnetic skyrmions are nontrivial topological magnetic textures that where predicted more than two decades ago.\cite{Bogdanov1989} Experimentally, they were first discovered in the form of a skyrmion lattice in 2009\cite{Muhlbauer2009a} and later also as isolated magnetic textures \cite{Romming2013}. Their presence has been observed in many device-relevant materials and their properties have been extensively summarized in several reviews~\cite{Nagaosa2013, Finocchio2016, Fert2017, Jiang2017}. Skyrmions are regarded as promising candidates for spintronic applications due to their mobility when driven by ultra-low currents\cite{Jonietz2010,Schulz2012} and their room temperature stability\cite{Yu2011b, Yu2012,Gilbert2015, Boulle2016, Woo2016, Tomasello2017a}. Particularly, the skyrmion racetrack memory has been a significant driver for intensively studying individual skyrmions.\cite{Fert2013, Tomasello2014, Zhang2015c, Muller2016}.

\begin{figure}
\includegraphics[width=1.0\textwidth]{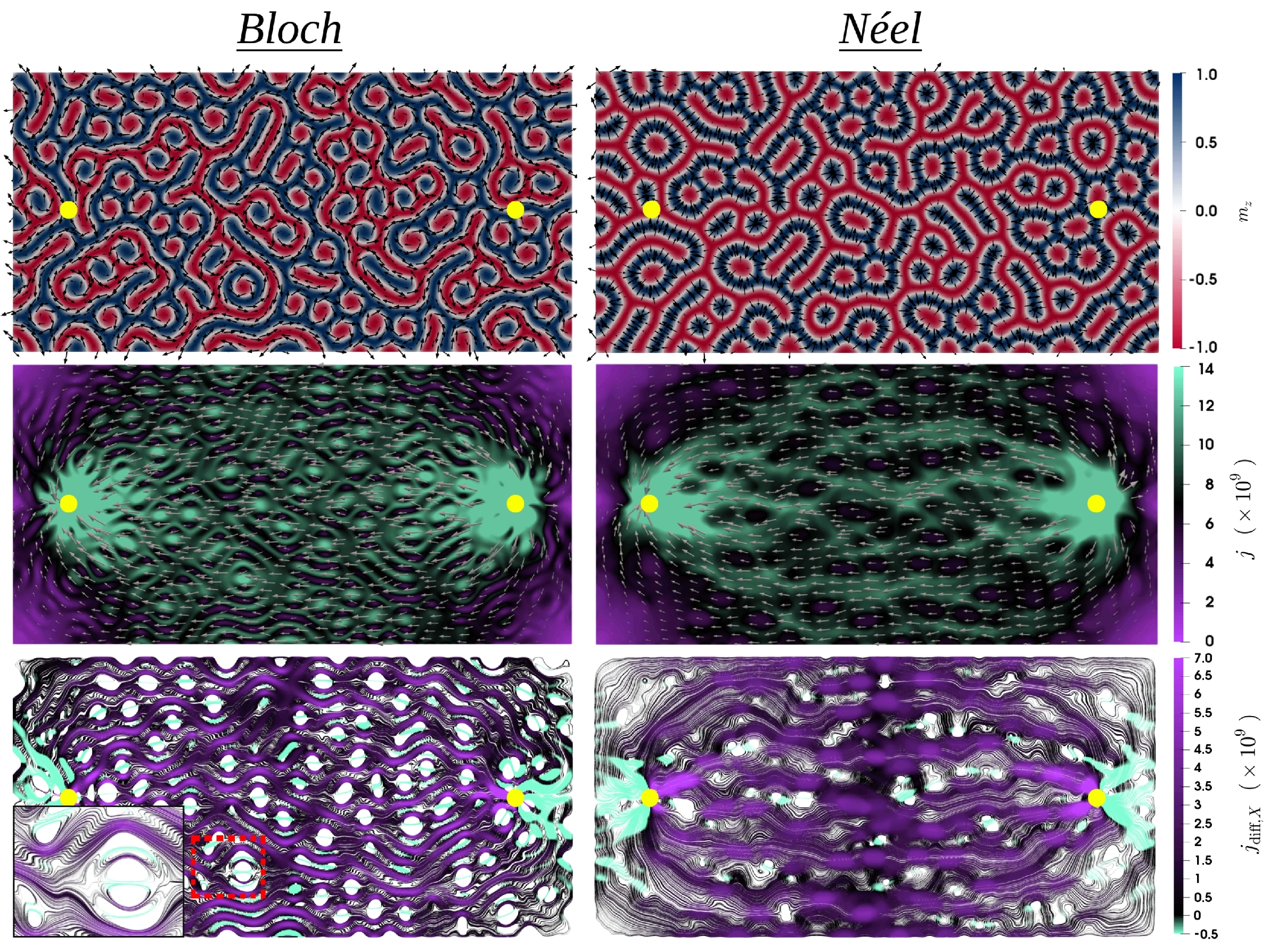}
\caption{Bloch (left column) and N\'eel (right column) skyrmion fabrics, where a voltage is applied in-between the contacts (yellow dots). 
Top row: Magnetization profiles. Color code and black arrows denote the out-of-plane and in-plane components respectively. 
Middle row: Current pathways through corresponding skyrmion fabrics. Color code and gray arrows denote current magnitude and direction respectively. 
Bottom row: Differential current flow lines showing the skyrmion-mediated AMR effects obtained by subtracting the trivial out-of-plane ferromagnetic current flow from that of the skyrmion fabric. Color code highlights regions where skyrmions enhance (green) and reduce (purple) flow along the negative $x$ direction revealing current backflows in the AMR-dominated regime. The inset shows a close-up view of the area enclosed by the red dashed rectangle.
The parameters used for these simulations are: $\alpha = 0.5$, 
$M_s=\SI{4.9e5}{\ampere\per\metre}$, 
$\sigma_0=\SI{5e6}{\siemens\per\metre}$, $a=1$,
$U = \SI{1e-3}{\volt}$, 
$A_{ex}=\SI{6e-12}{\joule\per\metre}$, 
$K_u=\SI{1.3e6}{\joule\per\cubic\metre}$, 
and $D_{B/N}=\SI{3e-3}{\joule\per\square\metre}$.
}
\label{fig:skyrmionfabrics}
\end{figure}

Not much attention has, however, been given to applications involving intermediate skyrmion phases known as ``skyrmion fabrics''\cite{Prychynenko2017}. These are phases that interpolate between single skyrmions, skyrmion crystals and magnetic domain walls\cite{You2015} (examples shown in the top panels of Fig.~\ref{fig:skyrmionfabrics}). In the past, skyrmion fabrics have been studied only to observe how the different phases contribute to transitions between them\cite{Milde2013}. 

We claim that skyrmion fabrics can provide a good basis for RC reservoirs in light of their random phase structure. The input signals can be realized via voltage patterns applied directly to the magnetic texture through various nanocontacts. Magnetoresisitive effects\cite{McGuire1975, Hanneken2015, Kubetzka2017} such as the anisotropic magnetorestance (AMR) will then guarantee that a certain magnetic texture will result in a unique corresponding current pattern throughout the reservoir (shown in middle panels of Fig.~\ref{fig:skyrmionfabrics}). 
The random skyrmion structure and corresponding current pattern will in turn model the reservoir node and weight structure discussed in the previous section.

The magnetization profiles and the current paths shown in Fig.~\ref{fig:skyrmionfabrics}
have been obtained by micromagnetic simulations using Micromagnum\cite{MicroMagnum} and selfwritten software extensions as in Ref.~\citenum{Prychynenko2017}, where the magnetization dynamics and the current paths have been computed selfconsistently.
The magnetization dynamics are given by the Landau-Lifshitz-Gilbert (LLG) equation for the magnetization direction  $\vect{m}= \vect M/M_s$  
with spin-transfer-torque effects:\cite{Berger1996,Slonczewski1996}
\begin{equation}
(\partial_{t} + \xi\, \vect j[U,\vect m] \cdot \vect{\nabla} ) \vect{m} = -\gamma \vect{m} \times \vect{B}_{\mathrm{eff}} +\alpha \vect{m} \times (\partial_{t} + \frac{\beta}{\alpha} \xi\, \vect j[U,\vect m] \cdot \vect{\nabla}) \vect{m}.
\end{equation}
Here, $M_s$ is the saturation magnetization, the factor $\xi =P \mu_{B}/(e M_{s})$ contains the polarization $P$ the electron charge $e$ and the Bohr magneton $\mu_B$. The effective magnetic field is given by $\vect{B}_{\mathrm{eff} }= -M_{s}^{-1} (\delta F[\vect{m}]/\delta \vect{m})$, where the micromagnetic free energy comprising exchange, anisotropy and dipolar interactions is:
\begin{equation}
F= \int \left( A_{\mathrm{ex}} (\nabla \vect{m})^{2} + K_{u} (1-m_z^2)
- \frac{\mu_0}{2} M_S \vect m \cdot \vect H_d(\vect m) \right) dV 
+F^{\mathrm{B}/\mathrm{N}}_{\mathrm{DMI}}[\vect m], 
\end{equation}
and 
$F^{\mathrm{B}}_{\mathrm{DMI}}= \int  D_B \vect{m}\cdot (\nabla \times \vect{m})\, dV$ describes Bloch and $F^{\mathrm{N}}_{\mathrm{DMI}}= \int D_N \vect{m} \cdot \lbrack(\mathbf{z}\times \nabla)\times \vect{m}\rbrack dV$ N\'eel DMI.\cite{Dzyaloshinsky1958, Moriya1960b, Thiaville2012} 
Note that the current density $j$ is a function of the applied voltage and the local magnetization. Since current density relaxation happens on a much faster time scale than the magnetization dynamics, it can be calculated self-consistently based on the AMR effect\cite{Kruger2011} through $\vect j[U, \vect m] = -\vect \sigma[\vect m] \cdot \vect E[U]$. Here the electric field induced by the applied voltage is calculated by solving the Poisson equation $\vect E=-\nabla \Phi$ with boundary conditions \mbox{$\Phi|_{c1}=-\Phi|_{c2}=U$} at the two contacts, and the conductivity tensor 
$\vect \sigma[\vect m] = \frac{1}{\rho_{\perp}} \mathbb{1} + \left( \frac{1}{\rho_{\parallel}} - \frac{1}{\rho_{\perp}} \right) \vect m \otimes \vect m$ varies with the local magnetization. We denote by $\rho_{\perp}$ ($\rho_{\parallel}$) the current resistivities for flows perpendicular (parallel) to the magnetization direction. 
Based on these we define $\sigma_0= (1/\rho_{\parallel} + 2/\rho_{\perp})/3$, and the AMR ratio $a=\frac{2(\rho_{\parallel}-\rho_{\perp})}{\rho_{\parallel} + \rho_{\perp}}$ as in Ref.~\onlinecite{Kruger2011, Prychynenko2017}. 

Previous work\cite{Prychynenko2017} has focused on details of current paths in the presence of single magnetic Bloch and N\'eel skrymions. As a result of the AMR effect, N\'eel skyrmions show a tendency to deflect current flow lines tangentially around them while Bloch skyrmions favor current flows through their centers. Furthermore, single skyrmions were shown to exhibit non-linear current-voltage characteristics due to the interplay of magnetoresitive effects and pinning. 

This work expands on previous results by exploring the effect of N\'eel and Bloch skyrmion fabrics on their resulting AMR-mediated current flow. In Fig.~\ref{fig:skyrmionfabrics}, we show examples of Bloch (left column) and N\'eel (right column) skyrmion fabrics. Their complex magnetic texture (top row) reflects on the total current flow between two contacts where a voltage difference is applied (middle row). To isolate the pure AMR effect of the skyrmion fabric we subtract the current response of a trivial out-of-plane ferromagnetic state and plot the resulting differential current flow (bottom row). We find that this artificially constructed AMR-dominated regime exhibits current backflows reminiscent of the recursive connectivity required in a RC network.

\begin{figure}
\includegraphics[width=1.0\textwidth]{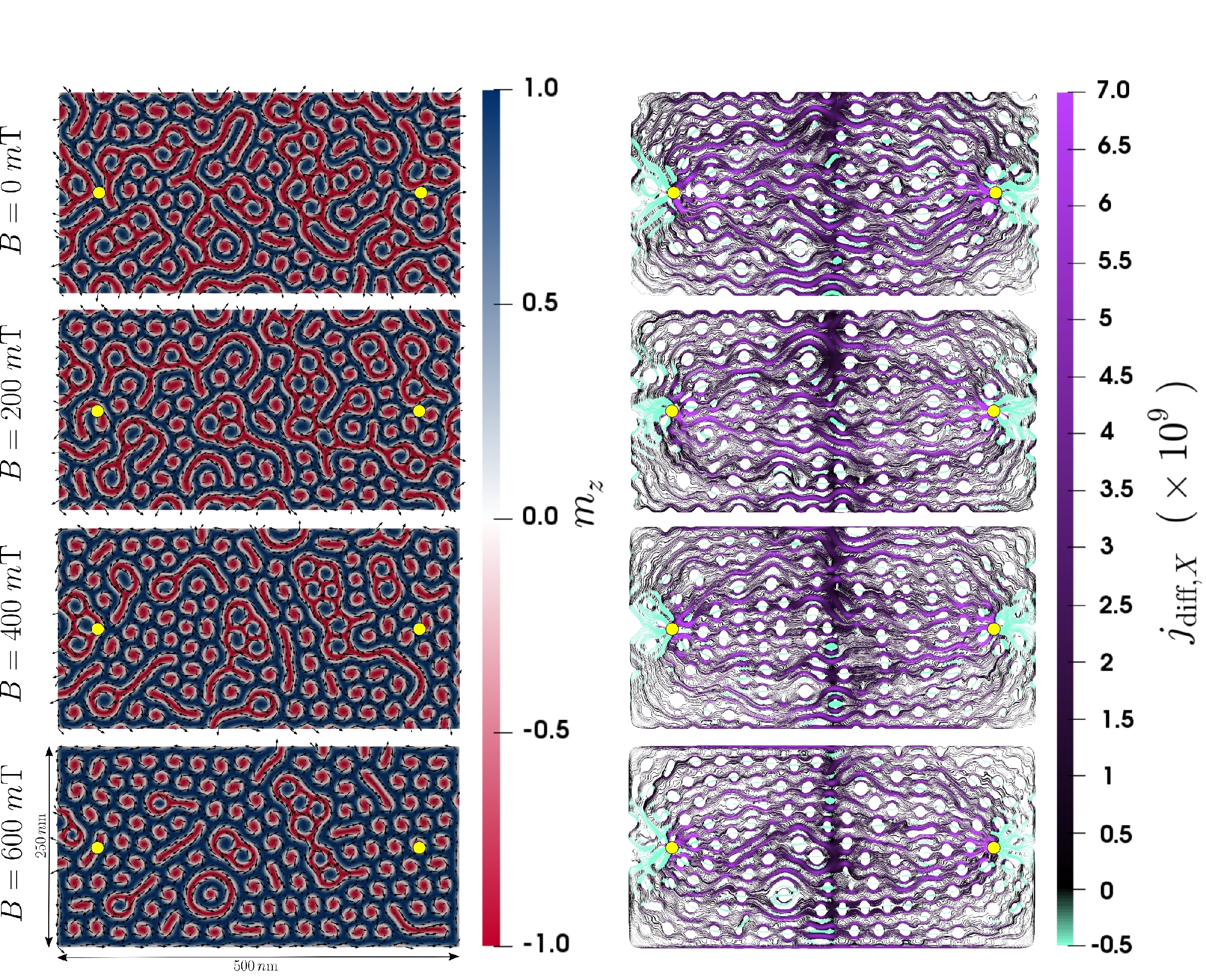}
\caption{
Magnetic texture as a function of applied out-of-plane magnetic field intensity (left column) and corresponding differential current flows (right column)}
\label{fig:skyrmionBfield}
\end{figure}

Due to the tunable properties of magnetic skyrmions, the skyrmion fabric can be tweaked by application of static magnetic fields. These alters the density of skyrmions thus tuning the effective node density throughout the sample. Figure~\ref{fig:skyrmionBfield} demonstrates this by showing how a Bloch magnetic texture and its resulting differential current flows both respond to variations of out-of-plane applied magnetic fields.

\section{Conclusions}
\label{sec:conclusions}

We have argued that skyrmion fabrics embedded in broken inversion symmetric magnetic substrates are potentially attractive options for implementing Echo State (ES) recognition and prediction systems. Leveraging results from previous work~\cite{Prychynenko2017} on nonlinear I-V characteristics of individual skyrmions, we show that skyrmion fabrics induce a strongly perturbed current flow through the magnetic texture as compared to one induced by a ferromagnetic state. By isolating these skyrmion-mediated effects, we have shown that the differential current flows exhibits regions of counter flow with recursive loops. 

The present calculation does not address the influence of thermal noise or potential grain structures in the magnetic material. Whereas our study is based solely on the anisotropic magneto-resistive (AMR) effect, other similar magnetization-modulated resistance effects could be included to tune and enhance our results.

The size scales of skyrmion fabrics are orders of magnitude smaller than other proposed implementations of reservoirs for ES networks, like memristor or optical networks. Furthermore, the characteristics of the skyrmion fabric (skyrmion density/size, domain wall width, etc...) can be altered by tuning the material properties and/or applying magnetic fields. These effectively tune both the dimensionality and net nonlinearity of the reservoir's dynamics, thus determining its performance.

\section*{Acknowledgments}

We acknowlegde discussions with Kai Litzius, Diana Prychynenko and Jairo Sinova.
G. B. appreciates support and useful discussions with Narayan Srinivasa of Intel.
We acknowledge the funding from the the German Research Foundation (DFG) under the Project No. EV 196/2-1, the Alexander von Humboldt Foundation and the Transregional Collaborative Research Center (SFB/TRR) 173 Spin+X.


%

\end{document}